\documentclass[aps,prl,twocolumn,showpacs,superscriptaddress]{revtex4-1}
\usepackage[utf8]{inputenc}
\usepackage{amsmath}
\usepackage{amsthm}
\usepackage{amssymb}
\usepackage{lmodern}
\usepackage{multirow}
\usepackage{booktabs}
\usepackage{graphicx}
\usepackage{subcaption}
\usepackage{placeins}
\usepackage[capitalise]{cleveref}
\usepackage{xcolor}
\usepackage{lineno}
\usepackage{ulem}

\begin{document}
%\linenumbers
\title{Use of a Generalized Energy Mover's Distance in the Search for Rare Phenomena at Colliders}
\author{M. Crispim Rom\~{a}o}
\email{mcromao@lip.pt}
\affiliation{LIP, Av. Professor Gama Pinto 2, 1649-003 Lisboa, Portugal}
\author{N. F. Castro}
\email{nuno.castro@fisica.uminho.pt}
\affiliation{LIP, Av. Professor Gama Pinto 2, 1649-003 Lisboa, Portugal}
\affiliation{Departamento de F\'{i}sica, Escola de Ci\^{e}ncias, Universidade do Minho, 4710-057 Braga, Portugal}
\author{J. G. Milhano}
\email{gmilhano@lip.pt}
\affiliation{LIP, Av. Professor Gama Pinto 2, 1649-003 Lisboa, Portugal}
\affiliation{Instituto Superior T\'{e}cnico, Universidade de Lisboa, Av. Rovisco Pais 1, 1609-001, Lisboa, Portugal}
\author{R. Pedro}
\email{rute@lip.pt}
\affiliation{LIP, Av. Professor Gama Pinto 2, 1649-003 Lisboa, Portugal}
\author{T. Vale}
\email{tiago.vale@cern.ch}
\affiliation{LIP, Av. Professor Gama Pinto 2, 1649-003 Lisboa, Portugal}
\affiliation{Departamento de F\'{i}sica, Escola de Ci\^{e}ncias, Universidade do Minho, 4710-057 Braga, Portugal}

\date{\today}

\begin{abstract}
    In this paper, we expand on the previously proposed concept of Energy Mover's Distance. The resulting observables are shown to provide a way of identifying rare processes in proton-proton collider experiments. It is shown that different processes are grouped together differently and that this can contribute to the improvement of experimental analyses. The $t\bar{t}Z$ production at the Large Hadron Collider is used as a benchmark to illustrate the applicability of the method. Furthermore, we study the use of these observables as new features which can be used in the training of Deep Neural Networks.
\end{abstract}

\maketitle

\section{Energy Mover's Distance as a Tool for Event Classification at Colliders}

One of the main goals of the analysis of experimental data in High Energy Physics is the classification of events, in an attempt to identify rare events of interest for different Physics studies. Such classification was a key aspect of the flagship results obtained from Large hadron Collider data, such as the Higgs boson discovery by the ATLAS and CMS experiments in 2012~\cite{ATLASHiggs,CMSHiggs}, the miriad of searches for new phenomena at colliders or, more recently, the observation of four top events~\cite{ATLAStttt} and the measurement of the $ttZ$ production cross-section~\cite{Aaboud:2019njj,CMS:2019too}.

In hadronic collisions, a very large number of particles are produced, which makes the classification of events particularly difficult. So, many of these classification tasks are relying more and more on complex Machine Learning classifiers, with the corresponding decision functions being difficult or impossible to interpret, and thus motivating the search for new interpretable physical observables. For such tasks, the complexity of final states can be treated by exploring not only the kinematic properties of the particles detected by the experiments, but also the overall flow of energy in an event~\cite{EventDistance}. That is the purpose of the recently proposed concept of a metric for the space of collider events based on the Energy Mover's Distance (EMD)~\cite{EventDistance,EnergyDistance}, where the similarity between two jets is quantified by computing the EMD between the distributions of the kinematics of each jet, providing a statistically-based intuition on how two jets are more or less similar.

In the current paper, we expand on the EMD definition by using global event properties, contributing to a better exploitation of the experimental information used in the search for rare events, which typically have cross-sections several orders of magnitude below the backgrounds affecting their measurement. In such cases, good discrimination between signal and background is a critical aspect to keep the experimental uncertainties under control and new variables contributing to correct classification of events can contribute to this goal~\cite{Mullin:2019mmh, Cesarotti:2020hwb}.

To generalise the EMD to full reconstructed events we introduce a new factor encoding information on the identity of the reconstructed physics objects present in each event. This generalised distance $d(I,J)$ between events $I$ and $J$ is then given by:
\begin{equation}
    \begin{aligned}
        d(I,J)={} & {\rm{\min_{f_{ij}}}} (\sum_{i,j}f_{ij}\Delta R_{ij}\times|p_{T,i}-p_{T,j}|\times ID(i,j)) \\
                  & + |E_I-E_J| \, ,
    \end{aligned}\label{eq:EMD}
\end{equation}
where the indices $i$ and $j$ run over the reconstructed final state objects in, respectively, events $I$ and $J$. $p_{T,i}$ ($p_{T,j}$) is the transverse momentum  \footnote{The transverse plane is defined with respect to the proton colliding beams.} of object $i$ ($j$) in event $I$ ($J$), $\Delta R_{ij}=\sqrt{\Delta \phi{_{ij}^2}+\Delta y{_{ij}^2}}$ is the rapidity-azimuth distance between objects $i$ (in event $I$) and $j$ (in event $J$), and $E_I$ ($E_J$) is the total reconstructed energy of event $I$ ($J$). The transport matrix elements $f_{ij}$, over which $d(I,J)$ is minimised, encode the optimal pairing between objects in events $I$ and $J$ and thus are such that:
\begin{equation*}
    \begin{aligned}
         & f_{ij} = \{0,1\},                          \\
         & \sum_j f_{ij} = 1, \ \sum_i f_{ij} = 1 \ ,
    \end{aligned}
\end{equation*}
where $f_{ij}=\{0,1\}$ is set explicitly to prevent, e.g., the energy of an electron in one event to be shared/associated to an electron and a muon on a second event to be considered as the optimal solution.

For each event, the reconstructed final state objects are the five leading small-$R$ jets, the two large-$R$ jets, the two leading electrons, the two leading muons and the missing transverse energy ($MET$). Jets are reconstructed from calorimeter energy clusters grouped using the jet finder algorithm anti-$k_t$~\cite{antikt} as implemented in the FastJet package~\cite{fastjet}, with radius parameter $R$=0.4 and $R$=0.8 for small- and large-$R$ jets, respectively. For events with fewer reconstructed objects, the absent ones are taken as null four-momentum vectors, providing a proxy to the object multiplicity in the event.

Before computing $d(I,J)$ from the simulated Monte Carlo samples, the events are first boosted to their centre-of-mass frame and then rotated in the tridimensional space to align the hardest object vertically in the $(y,\phi)$ plane. Since physical laws are Lorentz invariant, this procedure simply removes spurious information. Furthermore, as $d(I,J)$ are to encode a notion of similarity between the distributions of the kinematics of the objects between events, this procedure ensures that we are performing this comparison in a natural frame for each event.

The first term in \cref{eq:EMD} defines an overall distance between events weighted by the $p_T$ difference of their objects. The factor $ID(i,j)$ is introduced to encode information on the identity of the final state objects but implies that \cref{eq:EMD} cannot be, in general, interpreted as a distance in the geometric sense. While $d(I,J)$ is not a metric, it still provides the edges of a graph where each node is an event, i.e. the values $d(I,J)$ represent an adjancy matrix that can be used for network analyses, such as clustering which we will explore below. This is similar to the approach in~\cite{Mullin:2019mmh}. For simplicity we still call it a distance throughout the paper. $ID(i,j)$ consists of a variable scale factor that penalises the distance between two objects if they are of different type, where small-$R$ jets, large-$R$ jets, electrons, muons and $MET$ are considered different types of objects:
\begin{equation}
    ID(i,j) =
    \begin{cases}
        1          & \text{if } ID(i)=ID(j)                                    \\
        ID_{scale} & \text{if } ID(i)\neq ID(j) \text{ } (ID_{scale}\geq1) \ ,
    \end{cases}
\end{equation}

Computing the minimal distance implies minimizing the first term of the \cref{eq:EMD} with respect to the optimal transport, $f_{ij}$. We address this by using the Earth Mover's Distance algorithm implemented in the Python Optimal Transport {\texttt{ot}} library~\cite{optimalTransport}. Conceptually, the algorithm computes the minimal cost to transform one event into another. In practice this corresponds to finding $f_{ij}$ for which \cref{eq:EMD} is minimal.

The second term of the equation takes into account the total energy $E$ difference between the events $I$ and $J$. We study four variations of the distance between events, resulting from the combination of two options: adding the energy term or not and employing or not the $ID(i,j)$ scaler, i.e. setting $ID_{scale}=\{1,2\}$:
\begin{itemize}
    \item $d(I,J)$: $|E_I-E_J|$ not considered and $ID_{scale}=1$
    \item $d(I,J){^{ID}}$: $|E_I-E_J|$ not considered and $ID_{scale}=2$
    \item $d(I,J){_{\Delta E}}$: $|E_I-E_J|$ considered and $ID_{scale}=1$
    \item $d(I,J){_{\Delta E}^{ID}}$: $|E_I-E_J|$ considered and $ID_{scale}=2$
\end{itemize}

Despite which of the aforementioned options is at hand, distances between events of similar topology or kinematics will tend to be smaller while events yielding different final states will, in general, have larger distance values. This suggests that such an approach could help to differentiate between physical processes, providing an additional tool in tasks that demand high discrimination. Its impact could be especially relevant in studies of rare signals, often the case of searches for new physics, where the discriminative performance plays a crucial role. We highlight the adaptable nature of the constructed observables -- distances can be defined regardless of the event topology, the data filters employed or channel to be analysed -- and are therefore suitable for generic and model-independent searches for new physics and for anomaly detection.

In order to evaluate the impact of $ID_{scale}\neq 1$ and optimise this parameter, we also tested the values $ID_{scale}=\{1.5,3,10\}$. We found the resulting variations to be negligible with respect to the $ID_{scale}=\{1.5,2,3\}$ scan but we observe an improvement up to 20\% in the discriminant power of the distance calculated with $ID_{scale}=2$ when compared to $ID_{scale}=1$ for rare processes. Moreover, in the limit where $ID_{scale}=10$ and the weight of particle flavour in~\cref{eq:EMD} is an order of magnitude greater than kinematics, we obtain an improvement around 7\% in fake identification. Since a choice on the number of considered objects per event was made, we also tested the impact of such choice. This number can always be increased to adjust to specific physics scenarios with no harm to applications where fewer final objects are present, since an absent object enters as a null vector in the EMD calculation, therefore not contributing to the final distance. For instance, in the physics case explored in this paper, we observe no impact of increasing the number of small-$R$ jets to 10. Also, the possible overlap between large- and small-$R$ jets has no effect in our benchmark signal identification.

The time performance of the workflow is key to establish its practicability in a real experimental environment where billions of events need to be processed. In order to extract discriminative information about the events in a sample with $N$ events, we would, in principle, need to compute the distances between all the pairwise combinations of events in the sample, i.e. $N!/(2(N-2)!)$, which is not feasible even when resourcing to parallel computing and attaining an average processing time of around 1~ms/distance with the Python {\texttt{multiprocessing}} module. To overcome this drawback we define event references per process sample, that can later be used as the sample representatives to assess how far/near a given event is from the represented process. For a sample of 3k events, we compute the distances between all its events and then use a clustering technique to capture the structures existent in the data such as different kinematic regimes. We employ the k-Medoids clustering algorithm with the {\texttt{pyclustering}} Python library~\cite{pyclustering} and identify the medoid of each cluster, i.e., the central event according to \cref{eq:EMD}. The medoid approach was used in~\cite{EventDistance} to visualize sub-categories of jets. Here we expand this idea and use the medoids as the event references per process.

%\onecolumngrid

\begin{figure}
    \centering
    \caption{\label{fig:scatter_medoids} Distribution of (a) $t\bar tZ$ event clusters in the (jet $p_T$, $HT$) plane and (b) $t\bar tW$ events in the ($MET$,$HT$) plane, and respective medoids.}
    \begin{subfigure}[]{0.45\textwidth}\includegraphics[trim=0 0 0 25, clip, width=0.9\textwidth]{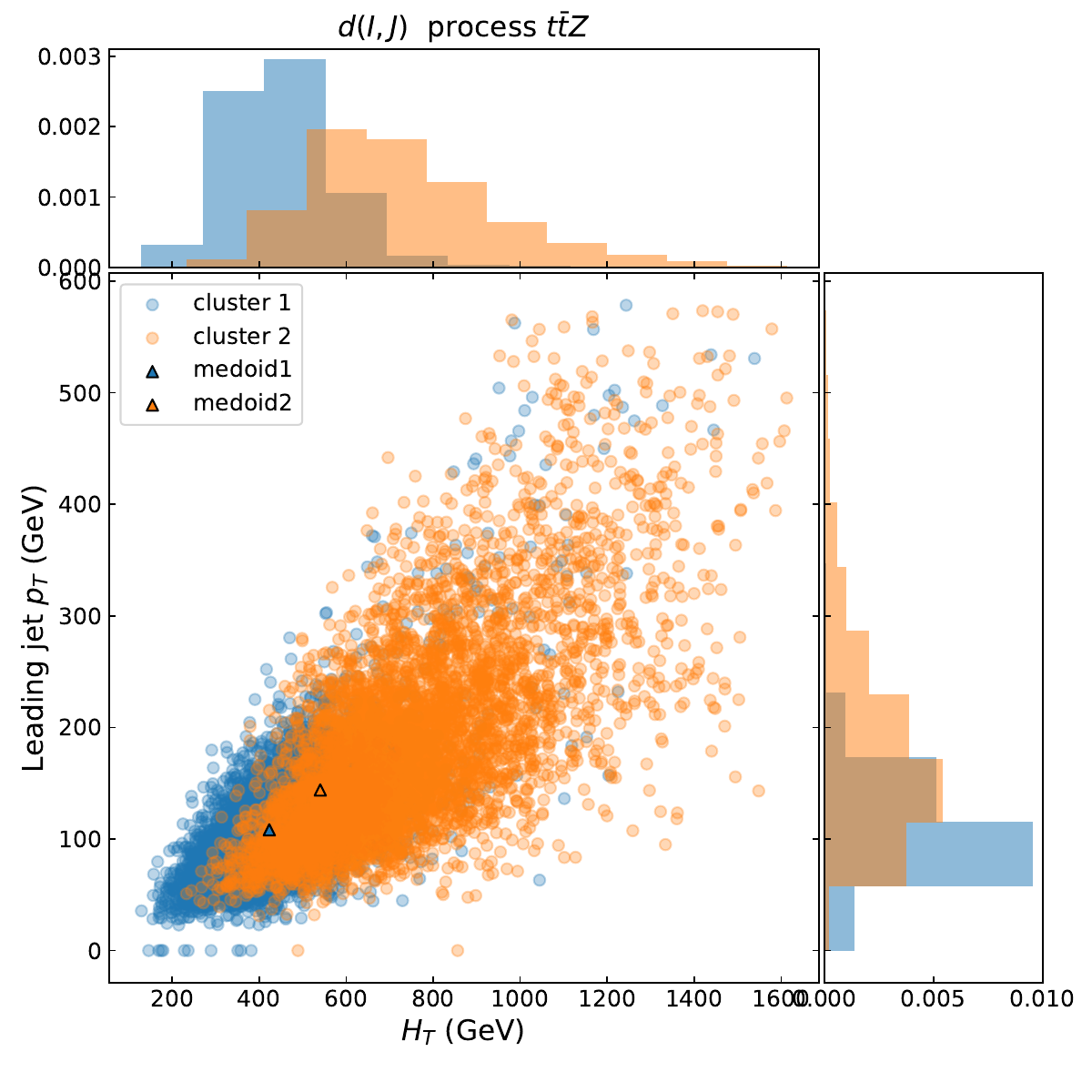} \caption{$t\bar tZ$ events} \end{subfigure}\quad\quad
    \begin{subfigure}[]{0.45\textwidth}\includegraphics[trim=0 0 0 25, clip, width=0.9\textwidth]{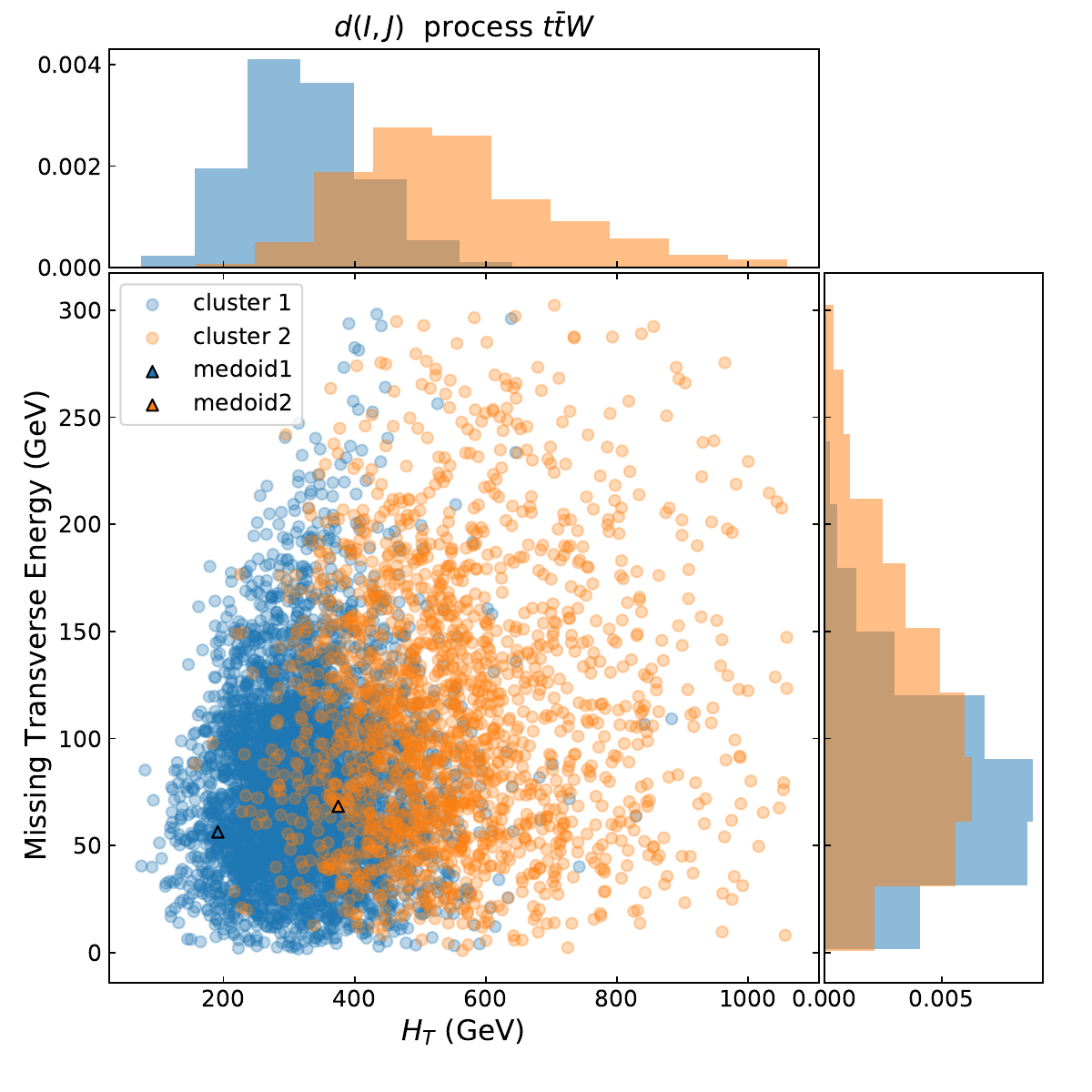} \caption{$t\bar tW$ events} \end{subfigure}
\end{figure}

%\twocolumngrid

\begin{figure}[htbp]
    \caption{\label{fig:medoids}Event Distances for a sample of 3~k $t\bar{t}Z$ simulated events (up) for all pairwise combination of events in the sample, (middle) for pairwise combination of events belonging to the same cluster and (bottom) between the cluster events and its medoid.}
    \includegraphics[width=0.4\textwidth]{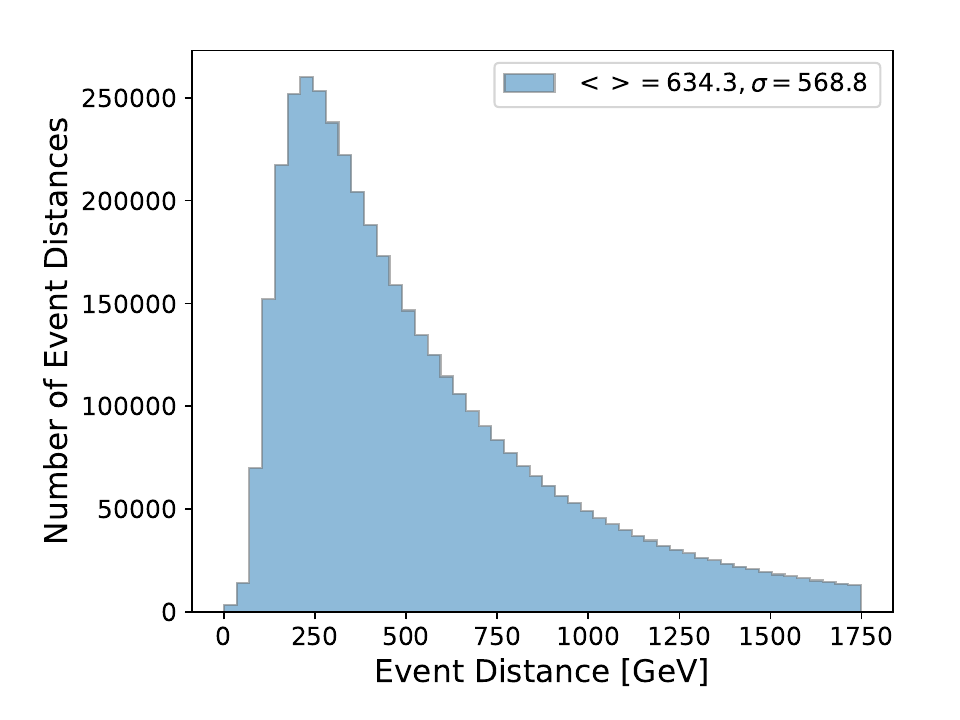}
    \includegraphics[width=0.4\textwidth]{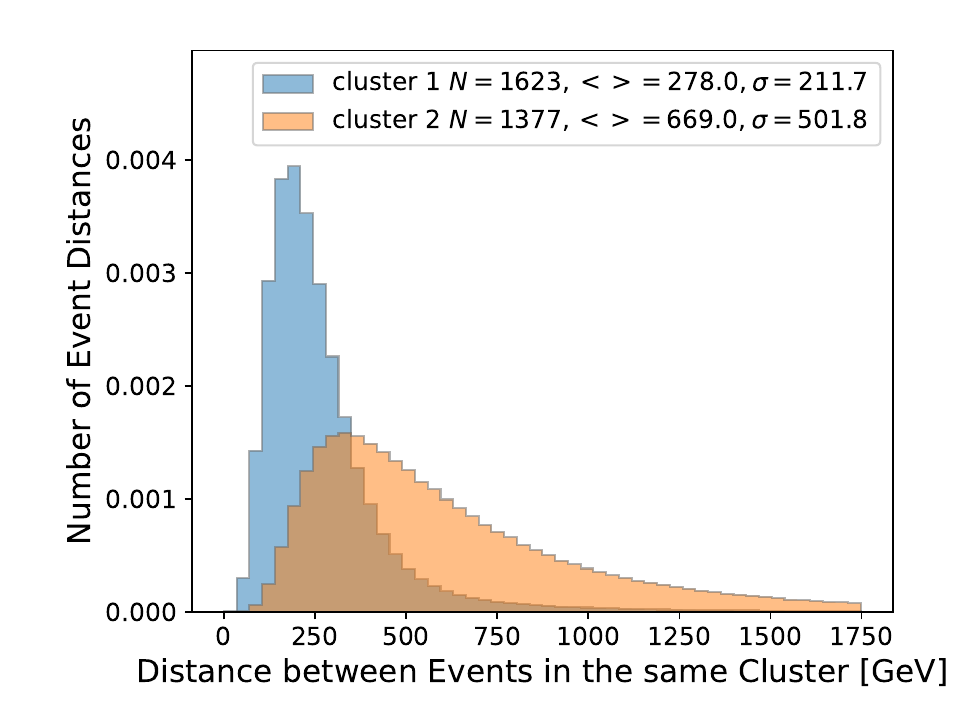}
    \includegraphics[width=0.4\textwidth]{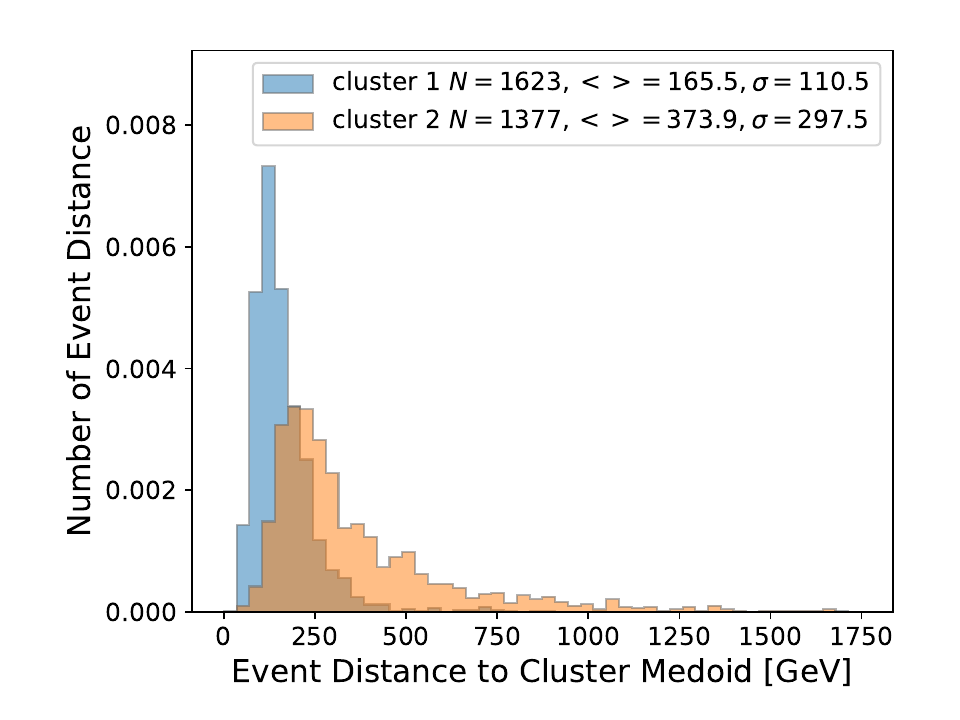}
\end{figure}

The number of clusters per sample is optimized using the Silhouette technique implemented in {\texttt{pyclustering}} ~\cite{pyclustering}. In this technique, the Silhouette score~\cite{rousseeuw1987silhouettes} -- which measures the cohesion of a cluster by contrasting the average distance between elements of the same cluster and their distances to the medoid -- is used to assert the optimal number of clusters by cutting-off when the average Silhouette scores in each cluster suffer a sharp drop if one would to add another cluster. Two clusters were found to be optimal.

Medoid events approximately correspond to the centre of clusters within the multidimensional event space. \cref{fig:scatter_medoids} illustrates this concept with bi-dimensional projections of $t\bar tZ$ and $t\bar tW$ event clusters and their respective medoids in the (jet $p_T$, $HT$) and ($MET$,$HT$) planes. $HT$ corresponds to the scalar $p_T$-sum of all the final state event objects. We can see that the medoids are approximately centred in the cluster bi-dimensional distributions.

\cref{fig:medoids} shows the distribution of the event distances for a sample of simulated $t\bar{t}Z$ events for all pairwise combination of events in the sample, for the pairwise combination of events belonging to the cluster, and between the cluster events and its medoid. Events within the first cluster are closer to each other as indicates the lower average and standard deviation. The second cluster is composed of events farther apart than in the first cluster but less scattered with respect to the original distribution, as seen from the lower standard deviation. The distances between the events and the cluster's medoids are even shorter as expected from the k-Medoids clustering. To understand how dependent these distributions and the subsequent results are on the initial 3k sample choice, we tested the use of statistically independent samples, also with 3k events, for all processes. These tests revealed no difference in the cluster distributions of $t\bar{t}Z$, showing that 3k events constitute a suitable statistical description of the process for the purpose of our study. Moreover, further results presented throughout the paper show no dependence on the initial sample.

\section{Physics case and Data Simulation}

We use simulated samples of proton-proton collision events generated with MADGRAPH5\_MCATNLO 2.6.5~\cite{madgraph} at leading order with a centre-of-mass energy of 13~TeV. The parton showering and hadronisation were performed with Pythia 8.240~\cite{pythia}, using the CMS underlying event tune CUETP8M1~\cite{CUETP8M1} and the NNPDF 2.3~\cite{NNPDF2.3} parton distribution functions. The detector simulation employs the Delphes 3.4.1~\cite{delphes} multipurpose detector simulator with the default configuration, corresponding to the parameters of the CMS detector.

The $t\bar{t}Z$ process is used as benchmark, corresponding to a typical measurement of a rare process at the Large Hadron Collider (LHC). Both the ATLAS and CMS Collaborations have considered trilepton final states for the measurement of the $t\bar{t}Z$ cross-sections~\cite{Aaboud:2019njj, CMS:2019too} and, therefore, we focus on such topologies. For this we select events with a final state composed of at least three leptons (\emph{i.e.} electrons or muons) compatible with the $Z\rightarrow \ell\ell$ decay and a leptonic top decay. Our main source of background is composed of $tX$ ($X={WZ,Zj}$), $t\bar{t}Y$ ($Y={W,Z,H}$) and dibosons ($WZ$ and $ZZ$). In addition, fake leptons arising from the misidentification of jets makes $t\bar{t}$+jets and $Z$+jets an additional non-negligible source of background.

In order to increase the efficiency of the trileptonic selection and obtain a good statistical representation of the different processes, the individual samples are generated with a dileptonic decay filter. Particle decays are implemented with MadSpin~\cite{madspin1,madspin2}, a simulator of narrow resonances decay that preserves spin and correctly implements its angular correlation scheme in the decay products.

Around 22~M events were simulated in order to achieve a statistical uncertainty which would be adequate to the analysis of 150~fb$^{-1}$ of data produced by the LHC:
\begin{itemize}
    \item 100~k for the $t\bar{t}Z$, $t\bar{t}W$ and $tX$ ($X={WZ,Zj}$) processes;
    \item 500~k for $t\bar{t}H$ and for each diboson ($WZ$ and $ZZ$) sample;
    \item 8~M for the $t\bar{t}$+jets process;
    \item 12~M for $Z$+jets events.
\end{itemize}
Each process was normalized to the expected yield for the considered benchmark luminosity of 150~fb$^{-1}$, assuming the Standard Model cross-sections computed at leading order with MADGRAPH5.

\section{EMD as High-Level Features}

In order to study the use of EMD as high-level features, we compute the distances between the events of all generated processes and the two medoids representing each process sample for each four distance options considered - $d(I,J)$, $d(I,J){^{ID}}$, $d(I,J){_{\Delta E}}$ and $d(I,J){_{\Delta E}^{ID}}$ - defined previously.

\begin{figure}[p!]
    \centering
    \caption{\label{fig:distances}Distribution of the Event Distances $d(I,J)$ to the (left) 2nd $t\bar{t}Z$ medoid and to the (right) 1st $WZ$ medoid for each process sample. All distributions are normalized to the unit area.}
    \includegraphics[trim={0 0 0 0.8cm},clip,width=0.5\textwidth]{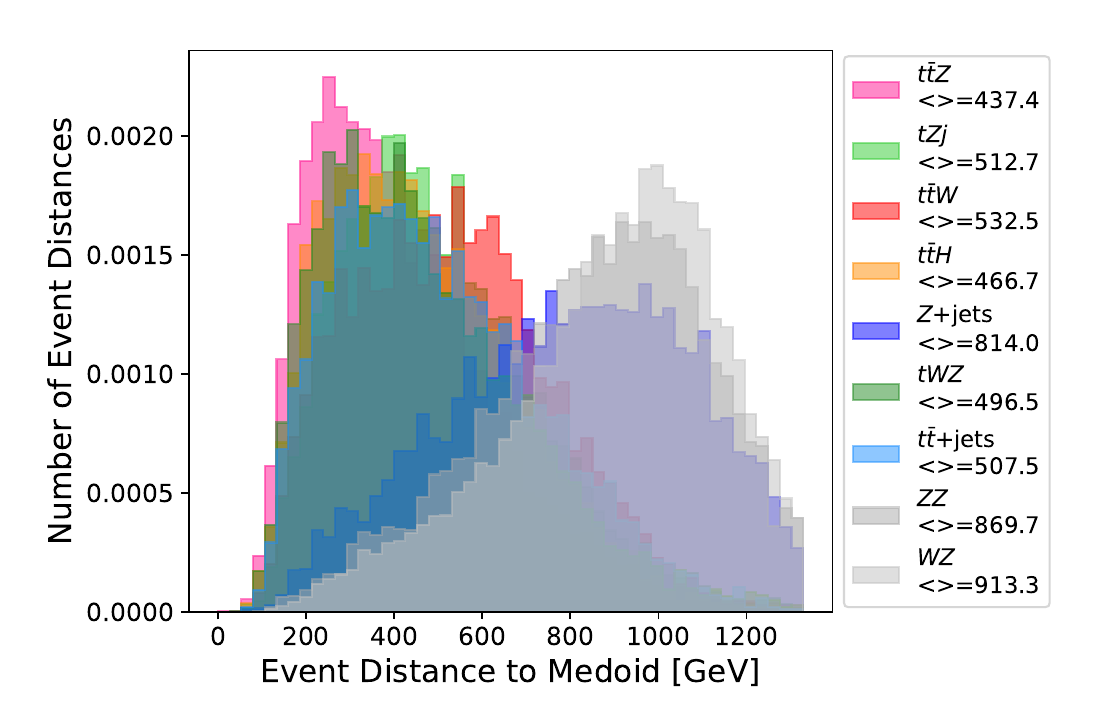}~\includegraphics[trim={0 0 0 0.8cm},clip,width=0.5\textwidth]{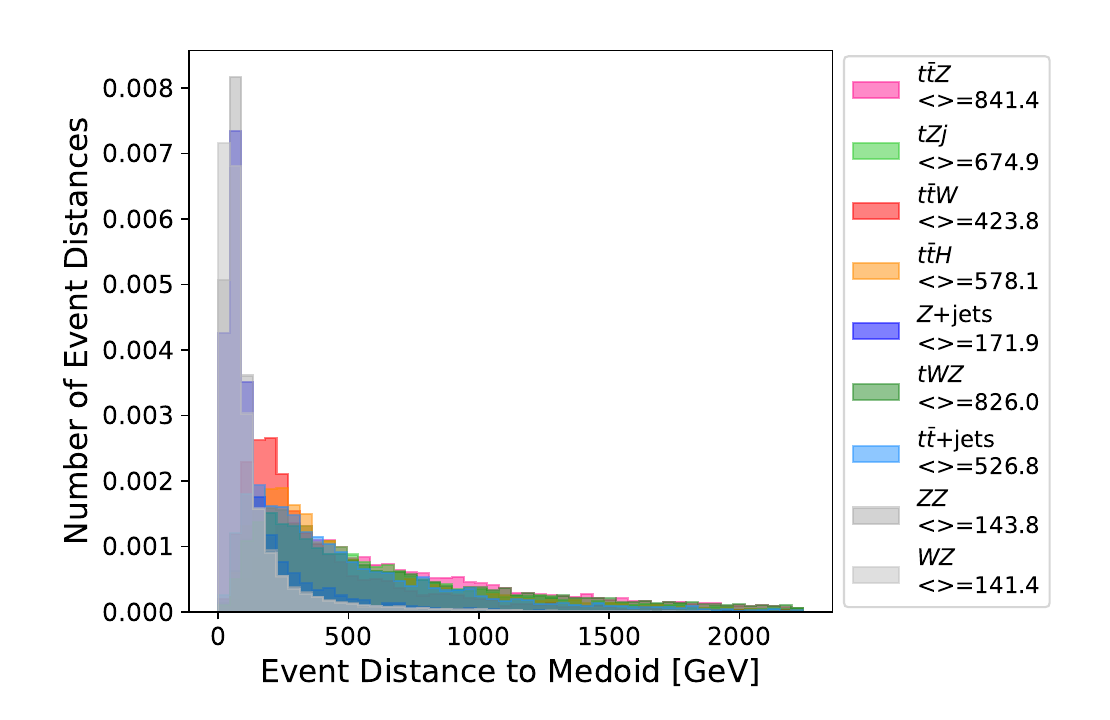}
\end{figure}

\cref{fig:distances} shows two example distributions of the event distances to a $t\bar{t}Z$ medoid and a $WZ$ medoid. Both the average and the median distance to the $t\bar{t}Z$ and $WZ$ medoids are lower for the $t\bar{t}Z$ and $WZ$ samples, respectively, as expected. Moreover, $ZZ$ and $Z$+jets events are in average close to the $WZ$ medoid, and the $t\bar{t}Y$ and $tX$ processes exhibit a short distance from the $t\bar{t}Z$ medoid. This observation provides evidence that the constructed set of distance observables has the ability to discriminate between event topologies. This conclusion is valid across all distributions of the distance observables and even if definite conclusions would require detailed detector simulation used by the LHC Collaborations~\cite{Abdullin:2011zz, Aad:2010ah}, the presented results look promising.

\onecolumngrid

\begin{figure}[htpb]
    \centering
    \caption{\label{fig:ROCdistances}Illustrative Receiver Operating Characteristic (ROC) curves, corresponding to the Event Distances to a medoid of (a) $t\bar{t}Z$, (b) $t\bar{t}H$, (c) $t\bar{t}W$, (d) $t\bar{t}$+jets, (e) $tZj$, (f) $tWZ$, (g) $WZ$, (h) $ZZ$ and (i) $Z+$jets, for each process sample. For each case, the ROC evaluates the task of distinguishing the process represented by the medoid reference from the remaining processes and the corresponding Area Under the Curve (AUC) is shown.}

    \begin{subfigure}[]{0.31\textwidth}\includegraphics[trim={0 0 0 0.8cm},clip,width=\textwidth]{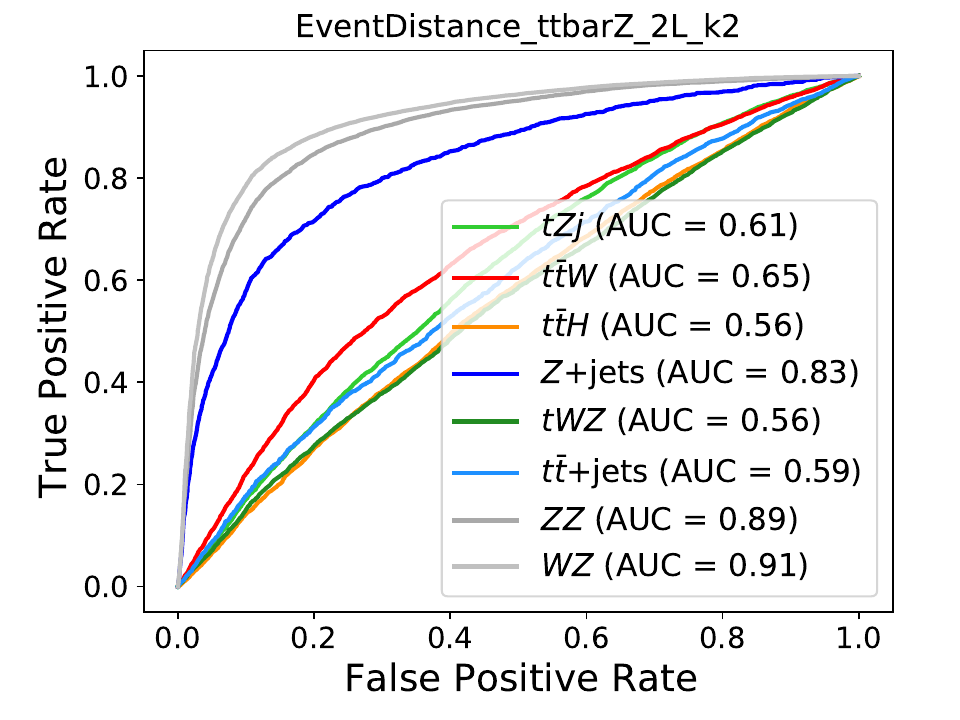}\caption{$d(I,J)$ to the 2nd $t\bar{t}Z$ medoid}\end{subfigure}
    \begin{subfigure}[]{0.31\textwidth}\includegraphics[trim={0 0 0 0.8cm},clip,width=\textwidth]{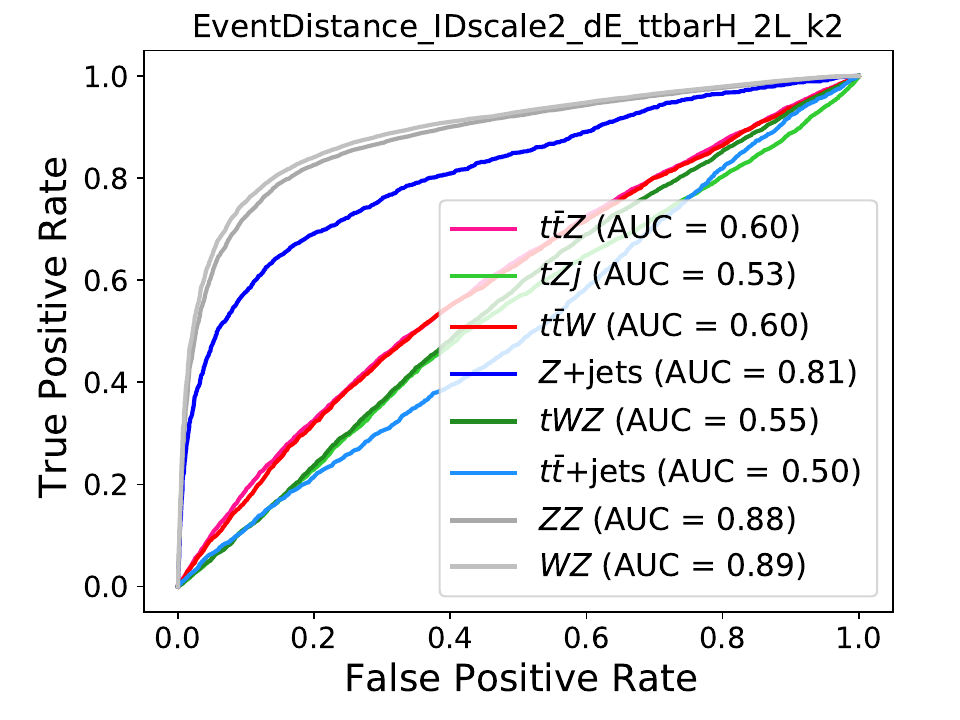}\caption{$d(I,J){_{\Delta E}^{ID}}$ to the 2nd $t\bar{t}H$ medoid}\end{subfigure}
    \begin{subfigure}[]{0.31\textwidth}\includegraphics[trim={0 0 0 0.8cm},clip,width=\textwidth]{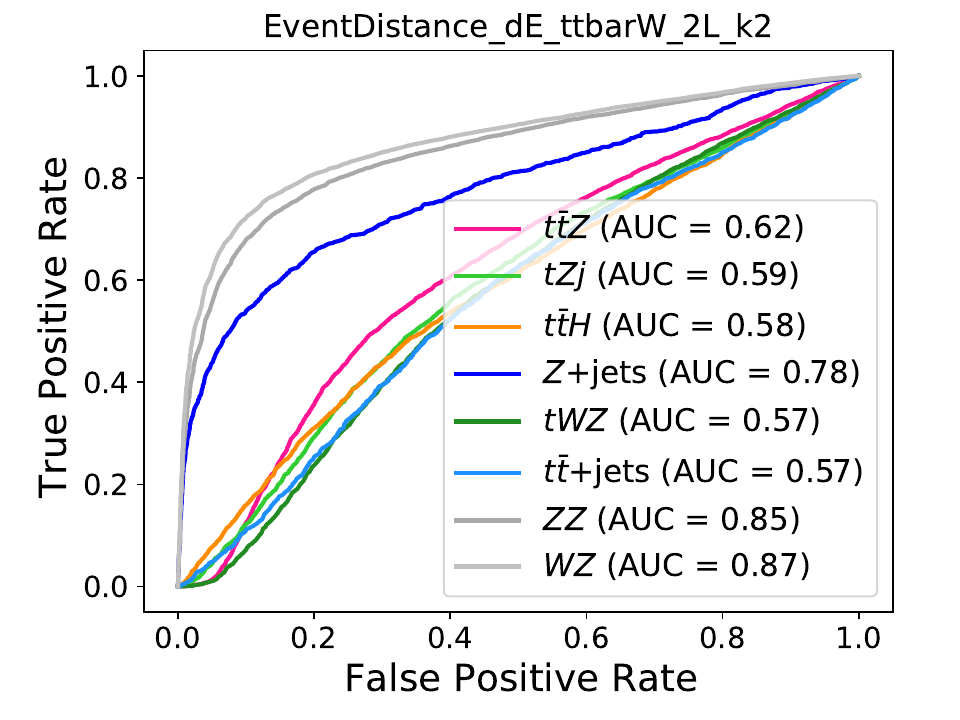}\caption{$d(I,J){_{\Delta E}}$ to the 2nd $t\bar{t}W$ medoid}\end{subfigure}\\
    \begin{subfigure}[]{0.31\textwidth}\includegraphics[trim={0 0 0 0.8cm},clip,width=\textwidth]{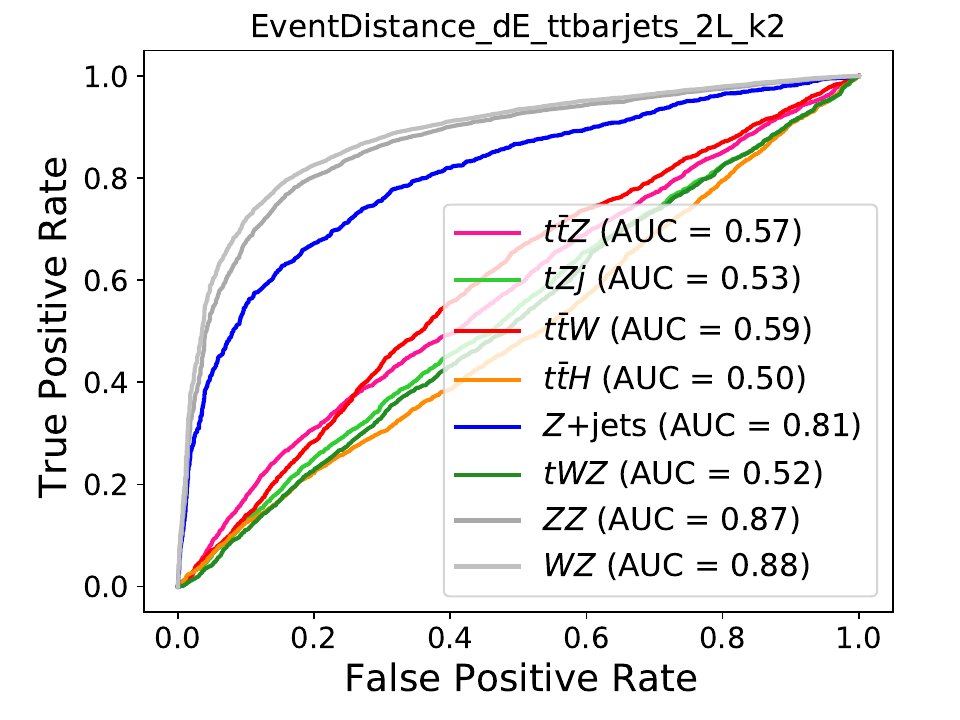}\caption{$d(I,J){_{\Delta E}}$ to the 2nd $t\bar{t}$+jets medoid}\end{subfigure}
    \begin{subfigure}[]{0.31\textwidth}\includegraphics[trim={0 0 0 0.8cm},clip,width=\textwidth]{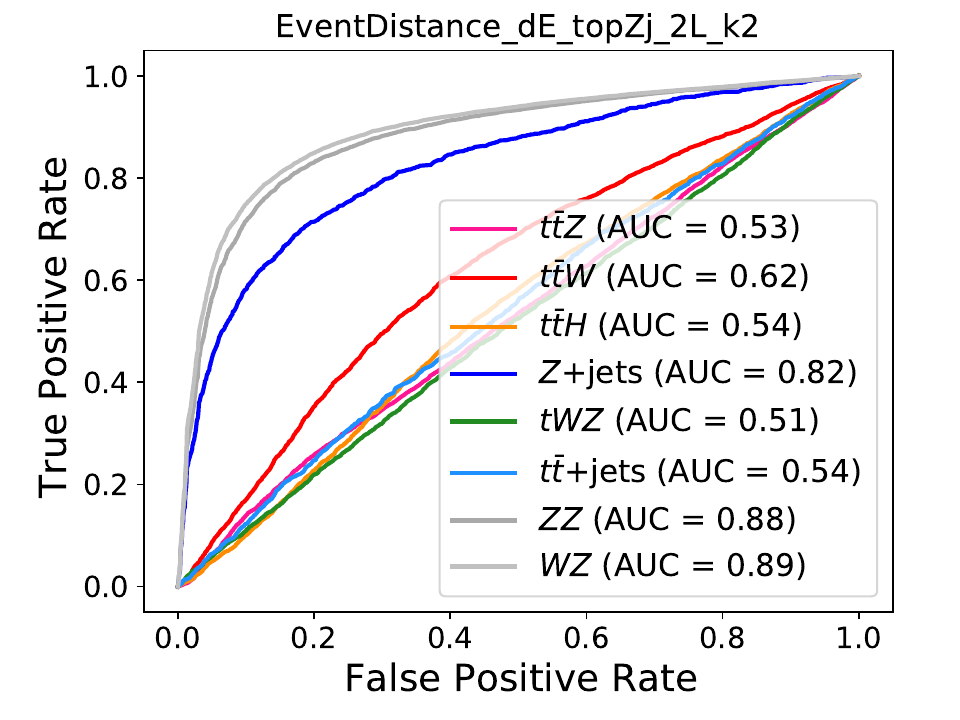}\caption{$d(I,J){_{\Delta E}}$ to the 2nd $tZj$ medoid}\end{subfigure}
    \begin{subfigure}[]{0.31\textwidth}\includegraphics[trim={0 0 0 0.8cm},clip,width=\textwidth]{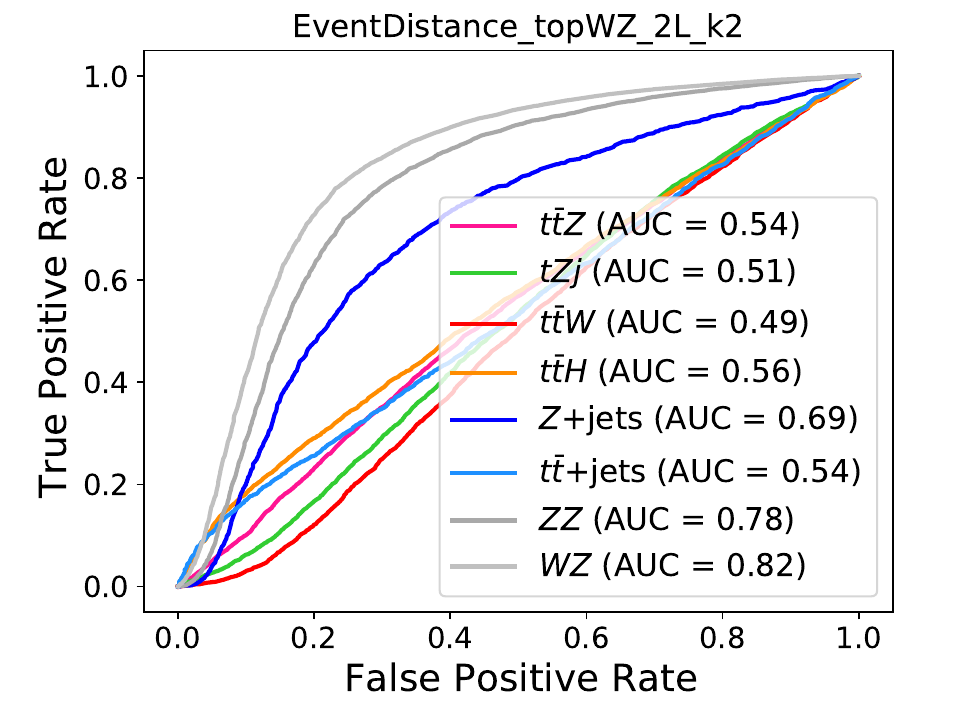}\caption{$d(I,J)$ to the 2nd $tWZ$ medoid}\end{subfigure}\\
    \begin{subfigure}[]{0.31\textwidth}\includegraphics[trim={0 0 0 0.8cm},clip,width=\textwidth]{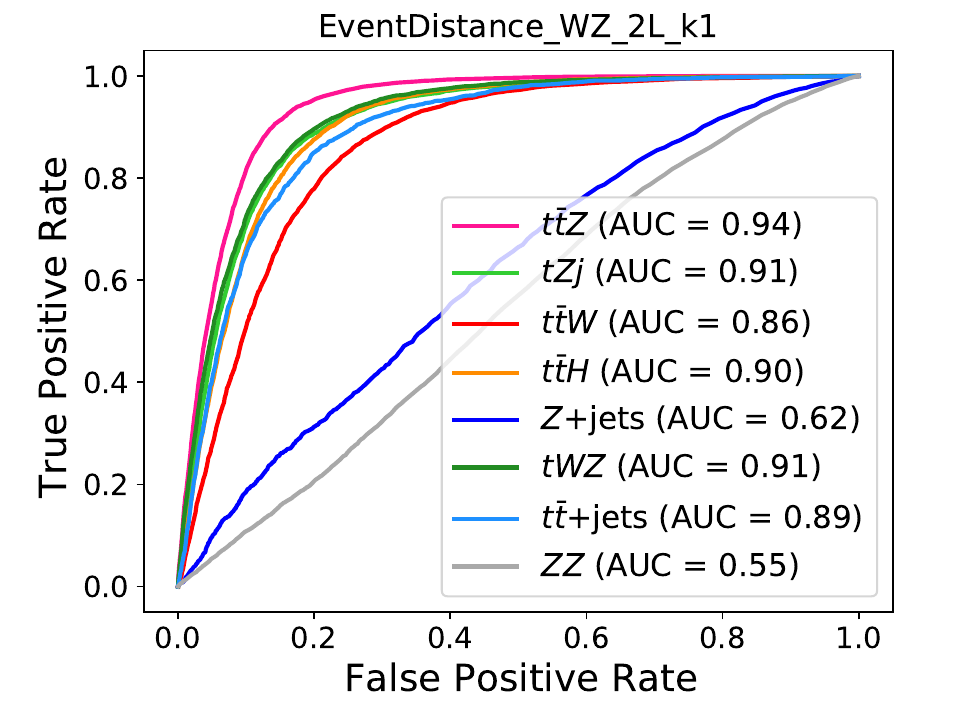}\caption{$d(I,J)$ to the 1st $WZ$ medoid}\end{subfigure}
    \begin{subfigure}[]{0.31\textwidth}\includegraphics[trim={0 0 0 0.8cm},clip,width=\textwidth]{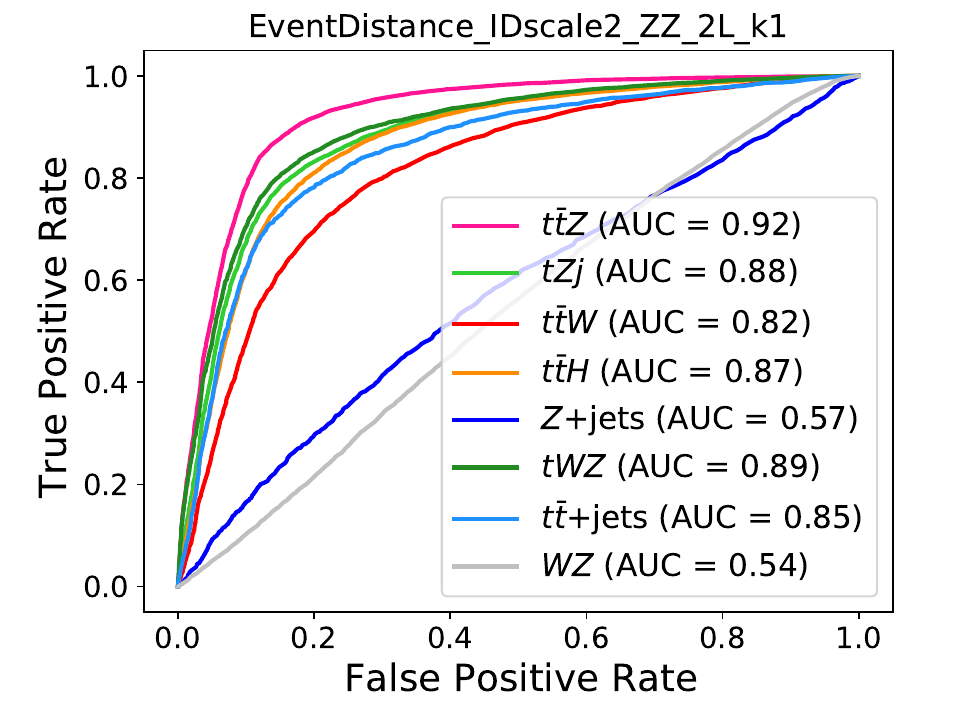}\caption{$d(I,J){^{ID}}$ to the 1st $ZZ$ medoid}\end{subfigure}
    \begin{subfigure}[]{0.31\textwidth}\includegraphics[trim={0 0 0 0.8cm},clip,width=\textwidth]{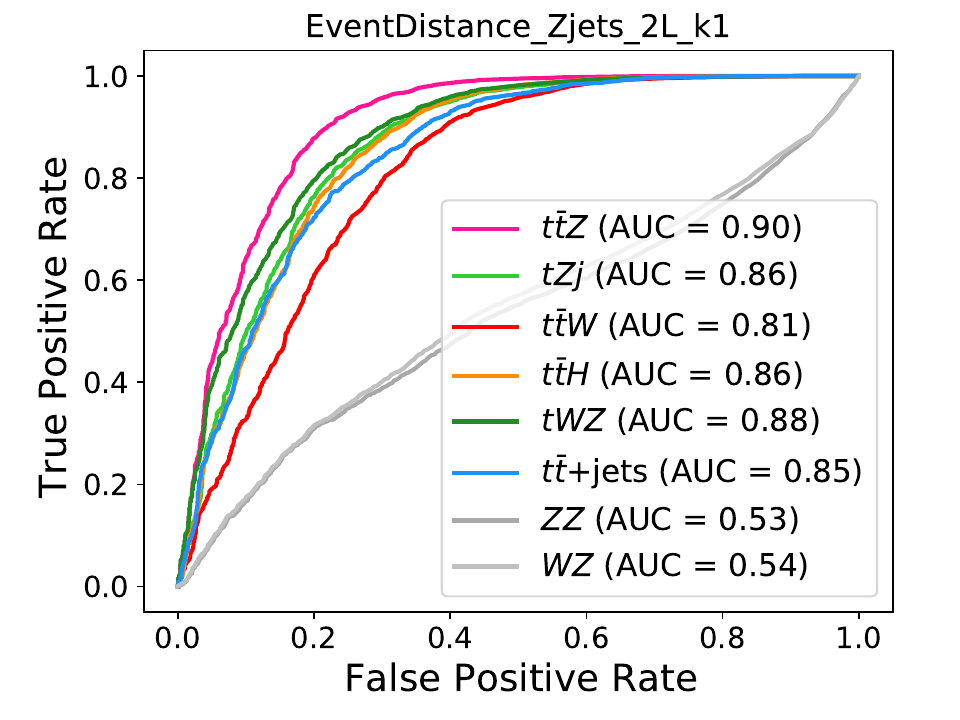}\caption{$d(I,J)$ to the 1st $Z$+jets medoid}\end{subfigure}\\
\end{figure}

\FloatBarrier

\twocolumngrid

The potential of the proposed generalization of EMD to distinguish physical processes is investigated by determining the distances of events with respect to each sample medoids and using it as a discriminant against the medoid event process. The corresponding Receiver Operating Characteristic (ROC) curves are shown in \cref{fig:ROCdistances} for one example medoid per process where average good performance is observed. Distances computed with respect to the $t\bar{t}Z$, $t\bar{t}Y$, $t\bar{t}$+jets and $tX$ medoids allow to discriminate the diboson and $Z$+jets processes. Conversely, distances to the diboson and $Z$+jets medoids are sensitive to processes containing top quarks. It is interesting to note that the constructed observable does not allow to distinguish $Z$+jets from diboson events. With the hardest jets originating from gluon splitting and the jet system recoiling against a dileptonic $Z$, the $Z+$jets events constitute indeed irreducible background against the diboson signals.

In order to further explore how this technique can be used in the context of High Energy Physics measurements we selected a set of high-level reconstructed event variables, from which we will derive a baseline to access its discriminant power, as well as to assess how different distances impact the corresponding separation performance. Following a typical choice of information set used in dedicated analysis at the LHC, the selected reconstructed variables used as features are:
\begin{itemize}
    \item ($p_T$, $\eta$, $\phi$) of the two leptons with the highest $p_T$;
    \item ($p_T$, $\eta$, $\phi$, $m$) of the two small-$R$ jets with highest $p_T$;
    \item $(b_1, b_2)$, being two binary variables indicating if the jets were tagged as originated by a $b$-quark;
    \item ($p_T$, $\eta$, $\phi$, $m$, $\tau_1$, ..., $\tau_5$) of the large-$R$ jet with the highest $p_T$;
    \item small-$R$ jet, electron, muon and large-$R$ multiplicities;
    \item scalar sum of all the reconstructed objects $p_T$, $H_T$;
    \item missing transverse energy and corresponding $\phi$;
\end{itemize}
with $\eta$ being the pseudo-rapidity of the corresponding object and $\tau_1$, ..., $\tau_5$ being the $N$-subjettiness observables of the large-$R$ jets~\cite{Thaler:2011gf,Stewart:2015waa}.

With both the event distances and the selected high-level features, we performed an exploratory analysis by embedding the events into a two-dimensional space using UMAP~\cite{mcinnes2018umap}, as implemented by~\cite{mcinnes2018umap-software}. For this, we standartized the features by subtracting their mean and divided by their standard deviation as to guarantee that all features are numerically of the same order of magnitude and adimensional. The embeddings for the selected features, for all the event distances, and for the combination of all event distances with the selected features can be seen in~\cref{fig:umap}. In this picture, we notice how, in a completely unsupervised manner, the embedding of the events through the selected features seems to be able to isolate clusters of events from different samples. The fact that the diboson events appear to be quite separated from those with a $t$-quark suggests that these events are the easiest to classify against the other classes, followed by $t\bar t Z$ events, which occupy mostly a single cluster. We also notice that fakes seem to mostly spread throughout all the clusters, highlighting the difficulty of isolating them. In the middle figure, we show the resulting embedding if we use all the event distances defined above. The 1st medoid of each sample, according to the $d(I,J)_{\Delta E}$ distance, is also represented in the embedding space and appears nearly centred on the sample distribution. Here again, we confirm the conclusion drawn in the previous section: these distances convey a notion of continuity from diboson events to $tX$ events. In the third figure, we used all the event distances in addition to the selected features. In this case, we notice that we can identify the same clusters as those appearing in the first picture, but that the event distances brought in the notion of continuity between events, continuously connecting some of the clusters.

\section{Deep Learning Application}

Since the event distances, either alone or combined with other high-level features, present a good discriminating power between physics processes, we went a step forward and studied how such discrimination compares with the one obtained through advanced machine learning techniques, namely Dense Neural Network (DNN). For this, we implemented DNN discriminants to perform the multiclassification task across the different sample classes (diboson, fakes, $tX$, $t\bar tY$ and $t\bar tZ$), corresponding to the physics process defined above.

We use TensorFlow 2.0~\cite{tensorflow2015-whitepaper} through its internal Keras API and followed the same sequential general architecture: input layer with width matching the number of input features, and a Softmax layer with five units as the output layer. The hyperparameters were fixed using HyperBand~\cite{li2017hyperband} as implemented by Keras-Tuner~\cite{kerastuner} for each set of features. The hyperparameteres tuned by the HyperBand algorithm were the number of layers (from 1 to 10), the number of units per layer (from 8 to 512 in steps of 8), dropout rate (from 0 to 0.5 in steps of 0.05), and the initial learning rate (from $10^{-5}$ to $10^{-2}$ over a log scale). We left fixed the activation function to LeakyReLu~\cite{xu2015empirical} and used Nadam optimizer~\cite{dozat2016incorporating}, and we used batch normalization~\cite{ioffe2015batch}. A 1:1:1 train-validation-test split was performed for the whole process and the final results presented here were derived from the test set.

\begin{figure}
    \centering
    \caption{\label{fig:umap} UMAP projections of three combination of features. The triangles show the 1st medoid of each sample according to the $d(I,J)_{\Delta E}$ distance. Units of final embedding space are arbitrary.}
%    \begin{subfigure}[]{0.28\textwidth}\includegraphics[trim=0 0 180 0, clip, width=\textwidth]{fig_04a.pdf}\caption{Selected Reconstructed Variables}\end{subfigure}
%    \begin{subfigure}[]{0.28\textwidth}\includegraphics[trim=0 0 180 0, clip, width=\textwidth]{fig_04b.pdf}\caption{All Event Distances \break} \end{subfigure}
%    \begin{subfigure}[]{0.40\textwidth}\includegraphics[width=\textwidth]{fig_04c.pdf}\caption{All Event Distances and Selected Reconstructed Variables}\end{subfigure}
     \includegraphics[width=\textwidth]{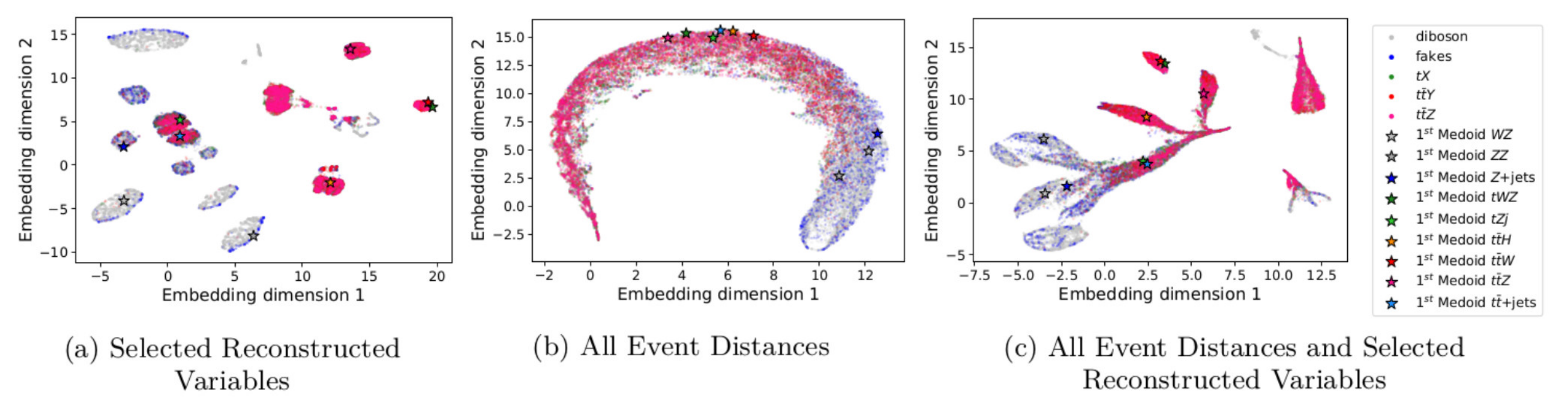}
\end{figure}

\begin{figure}
    \centering
    \caption{\label{fig:NormalisedCM} Normalised Confusion Matrices for all DNN, depending on the trained features.}
    \begin{subfigure}[]{0.32\textwidth}\includegraphics[trim={2.9cm 0 0.8cm 0.86cm}, clip,width=\textwidth]{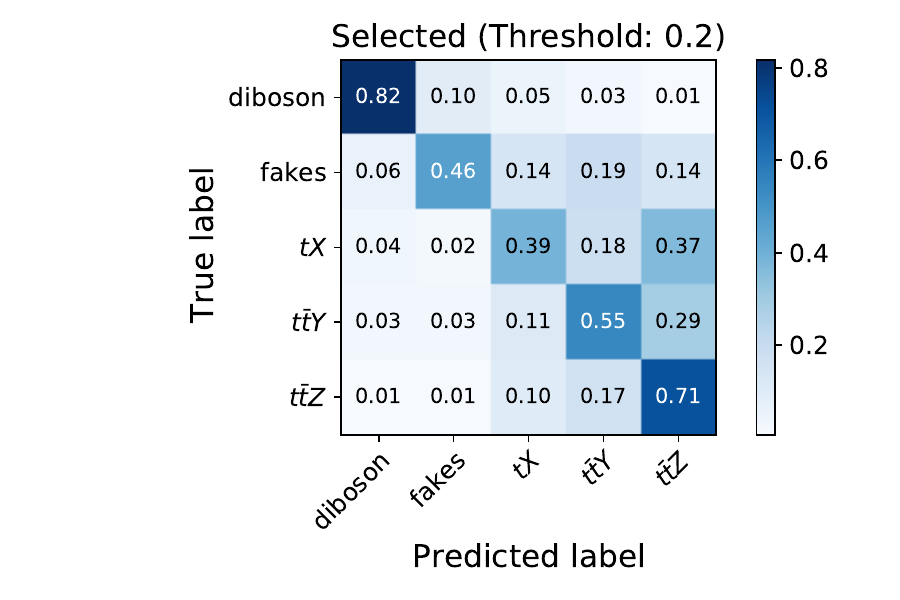}\caption{Selected Reconstructed Variables. Operating point of 0.2 \break}\end{subfigure}
    \begin{subfigure}[]{0.32\textwidth}\includegraphics[trim={2.9cm 0 0.8cm 0.86cm}, clip,width=\textwidth]{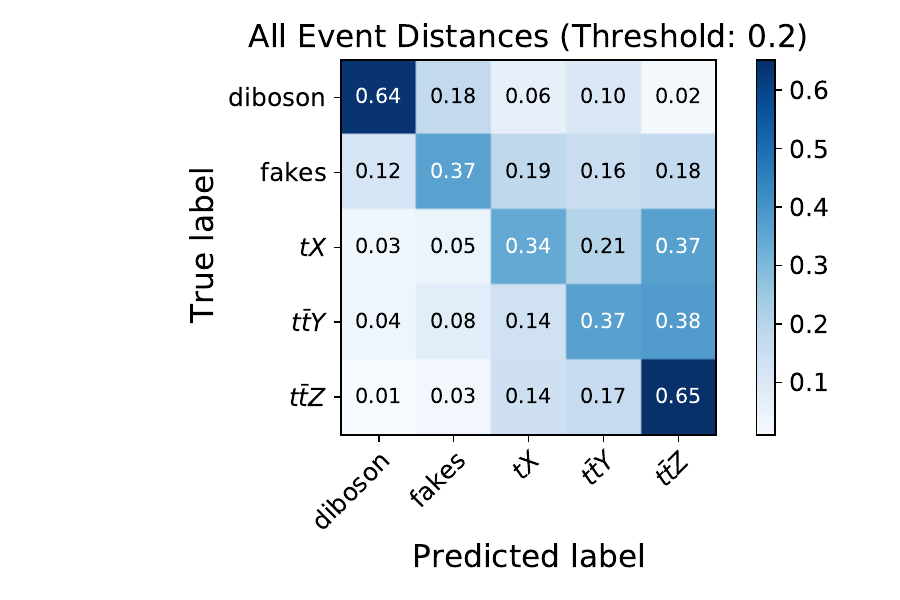}\caption{All Event Distances. Operating point of 0.2 \break}\end{subfigure}
    \begin{subfigure}[]{0.32\textwidth}\includegraphics[trim={2.9cm 0 0.8cm 0.86cm}, clip,width=\textwidth]{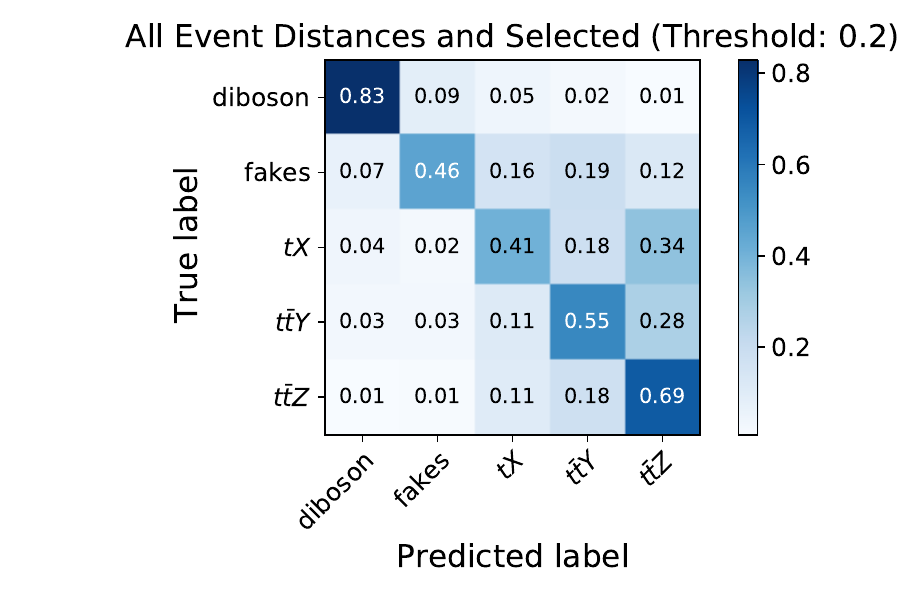}\caption{All Event Distances and Selected Reconstructed Variables. Operating point of 0.2}\end{subfigure} \\
    \begin{subfigure}[]{0.32\textwidth}\includegraphics[trim={2.9cm 0 0.8cm 0.86cm}, clip,width=\textwidth]{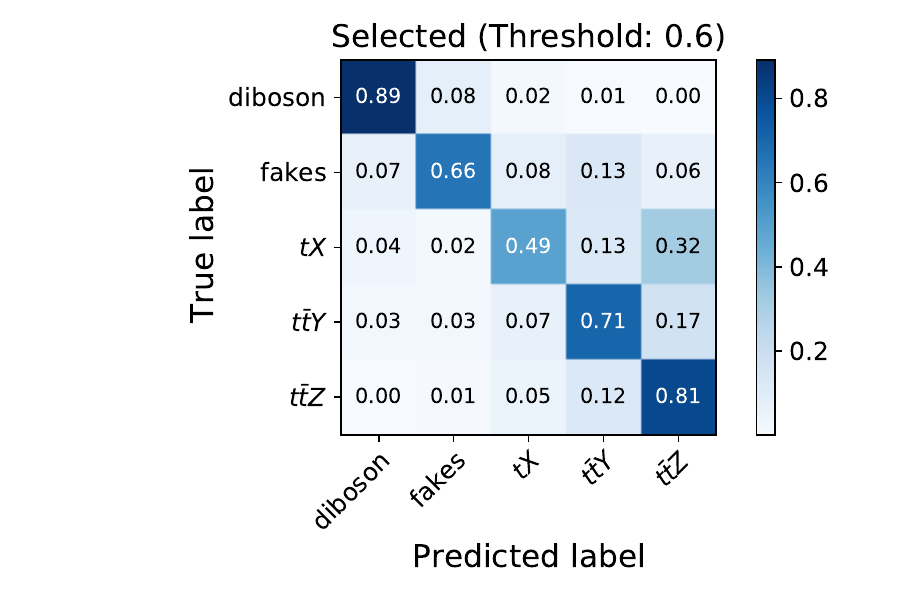}\caption{Selected Reconstructed Variables. Operating point of 0.6 \break}\end{subfigure}
    \begin{subfigure}[]{0.32\textwidth}\includegraphics[trim={2.9cm 0 0.8cm 0.86cm}, clip,width=\textwidth]{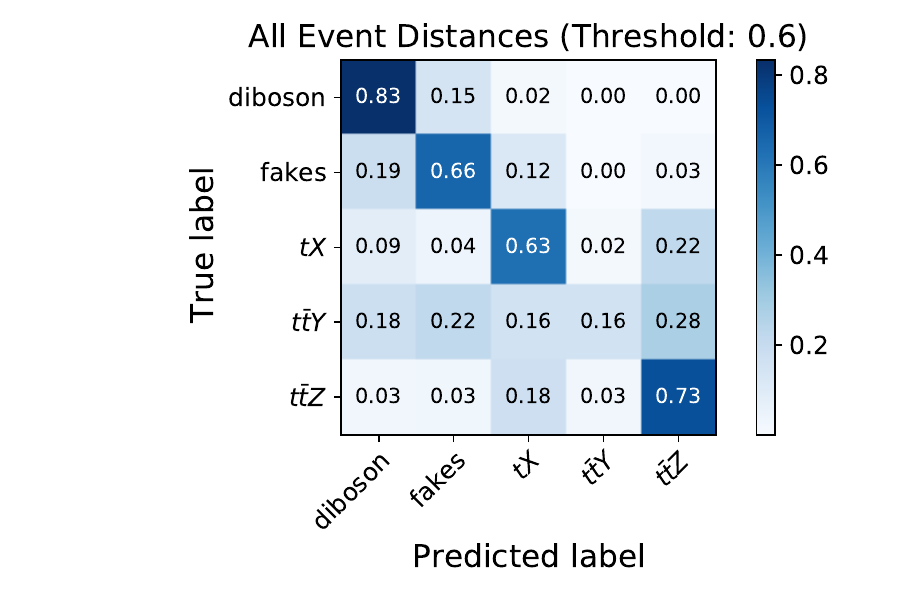}\caption{All Event Distances. Operating point of 0.6 \break}\end{subfigure}
    \begin{subfigure}[]{0.32\textwidth}\includegraphics[trim={2.9cm 0 0.8cm 0.86cm}, clip,width=\textwidth]{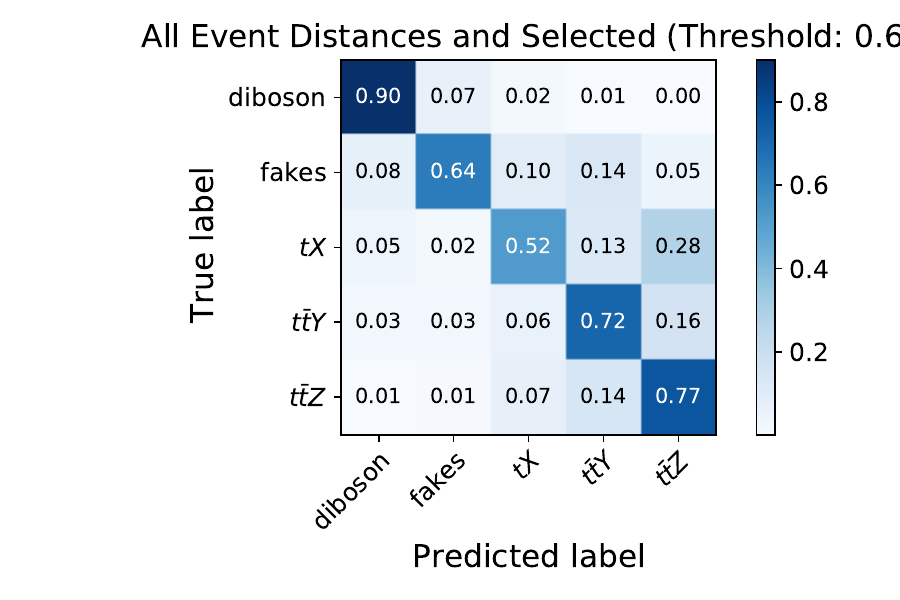}\caption{All Event Distances and Selected Reconstructed Variables. Operating point of 0.6}\end{subfigure}
\end{figure}

In~\cref{fig:NormalisedCM} we show the confusion matrices for the three combinations of features of \cref{fig:umap}, for two operating points. The first operating point (up) is defined by only accepting predictions, where the most likely prediction is greater than 0.2. This excludes the cases where the DNN cannot differentiate between any class and predicts 0.2 for all five classes. The second operating point (down) is set to 0.6, which will only retain more confident predictions. In these confusion matrices we notice that for low operation points, the inclusion of event distances to the high-level features has little performance impact. For a high operating point, we see that the event distances seem to help retain a fair discriminative power of fakes and improve $tX$ identification. These operating points are only meant to illustrate the potential of the proposed method since for each specific analysis they would need to be optimized. A more realistic experimental analysis would also need to take into account the effect of systematic sources of uncertainty in such optimization.

\begin{figure}
    \centering
    \caption{\label{fig:roc_auc_dnn}Areas under the ROC curves for all signals for all feature combinations. The left figure shows the obtained values for the areas under the ROC curves while in the right one the values were normalized to the values obtained for the baseline of selected reconstructed features (last row).}
    \begin{subfigure}[]{0.48\textwidth}\includegraphics[trim={0.7cm 0 3cm 0}, clip, width=\textwidth]{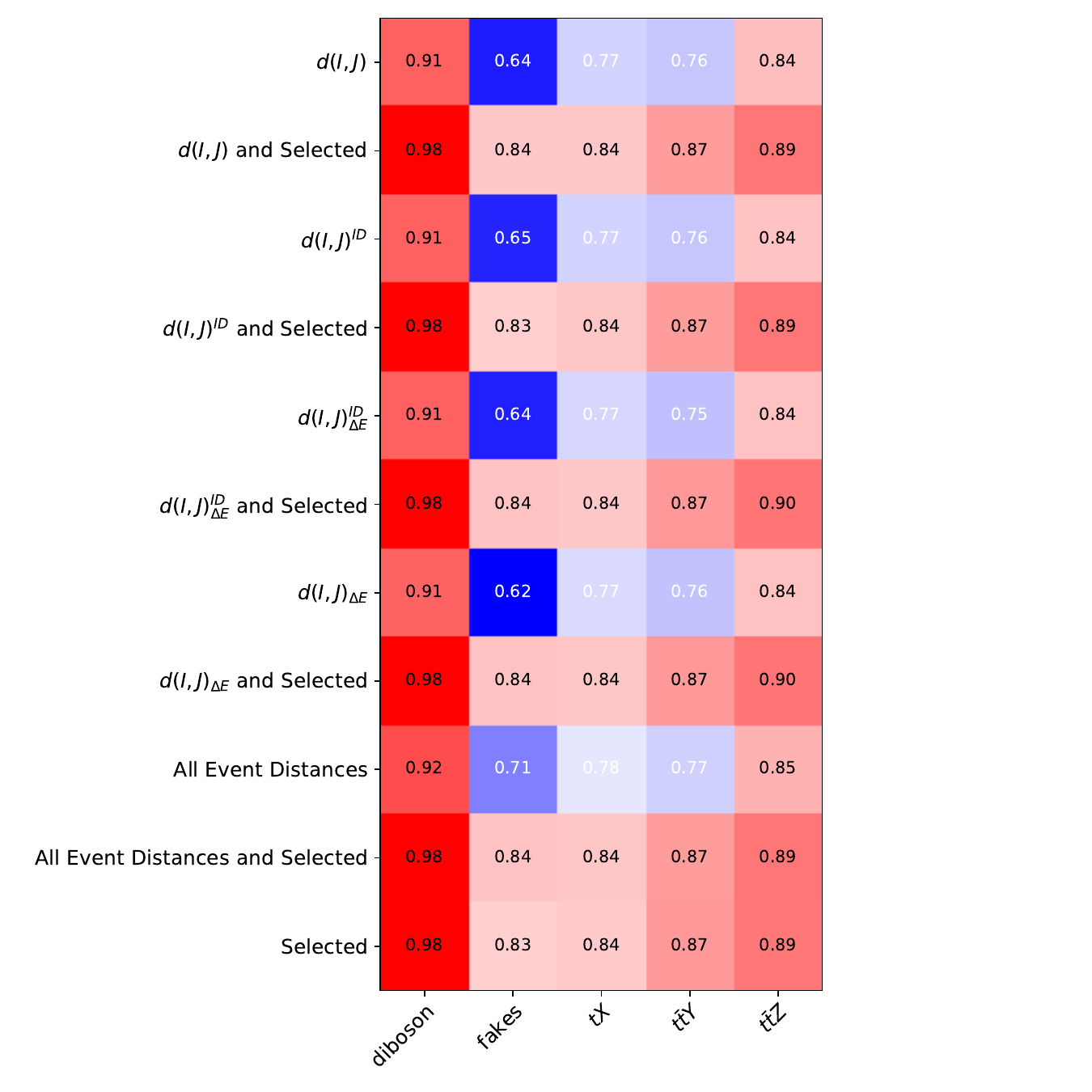}\end{subfigure}
    \begin{subfigure}[]{0.48\textwidth}\includegraphics[trim={0.7cm 0 3cm 0}, clip, width=\textwidth]{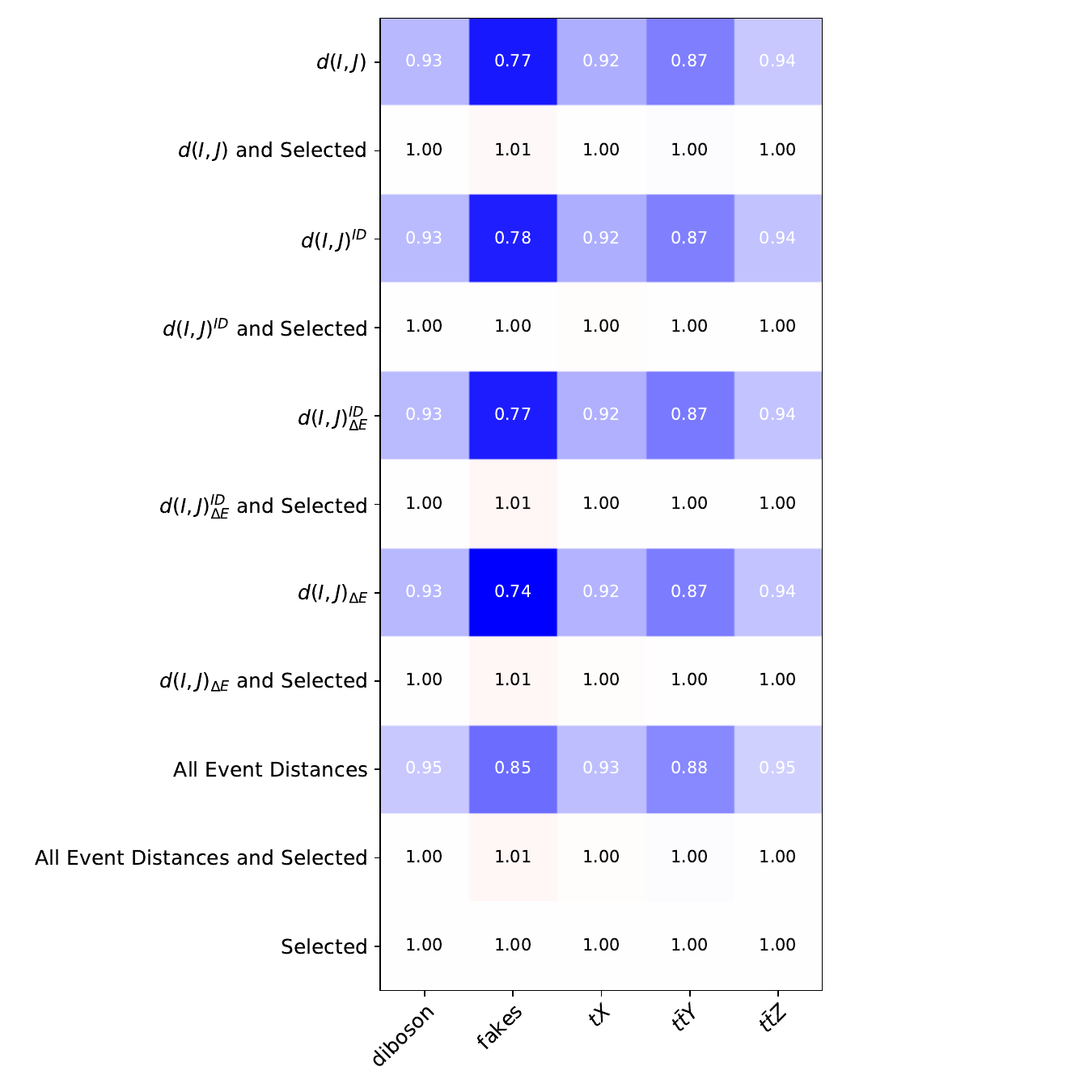}\end{subfigure}
\end{figure}

Next, in~\cref{fig:roc_auc_dnn}, we present the values of the areas under the ROCs for the multiclass discrimination using the different feature combinations on top, and how these compare to the baseline of using the selected reconstructed variables when training a DNN below. We see that each distance has discriminative power which depends slightly on the class we are trying to isolate and the set of selected features encloses a better discriminant power than the distance observables. The latter is expected since the selected features contain information, such as $b-$tagging, which is not present on the distance observables. Despite of this, the distance variables reach a competitive performance for diboson and $t\bar tZ$ identification. In addition, for example, the task of identifying fakes seems to benefit from the inclusion of the distances that include $\Delta E$ contribution to the distance, and even more from taking all distances into account. Identifying the remainder of the classes seem to benefit little or not at all from the inclusion of different event distances as features.

\begin{figure}
    \centering
    \caption{\label{fig:roc_fakes}ROC curves for the (left) Fake and (right) $t\bar tZ$ identification under different combination of features.}
    \includegraphics[trim={0 0 0 1cm}, clip,width=0.48\textwidth]{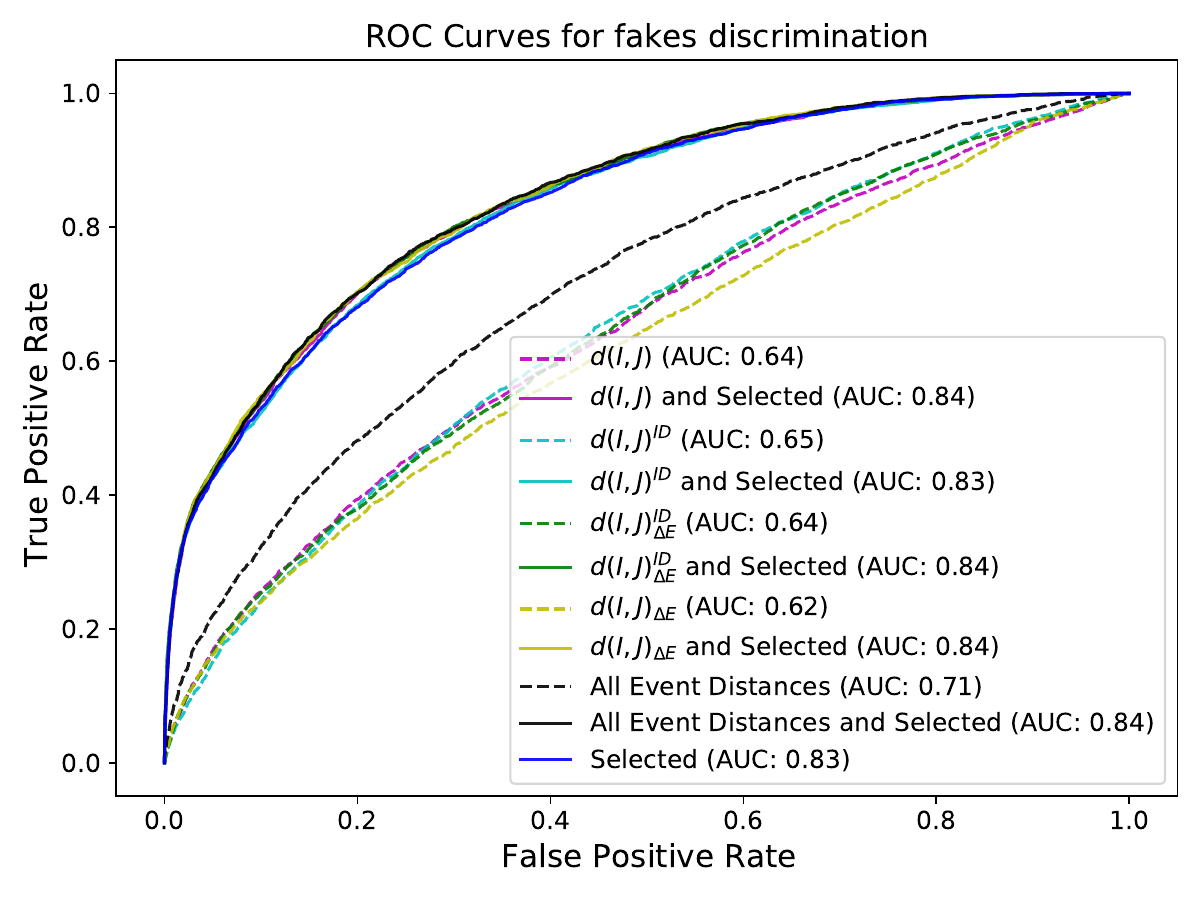}
    \includegraphics[trim={0 0 0 1cm}, clip,width=0.48\textwidth]{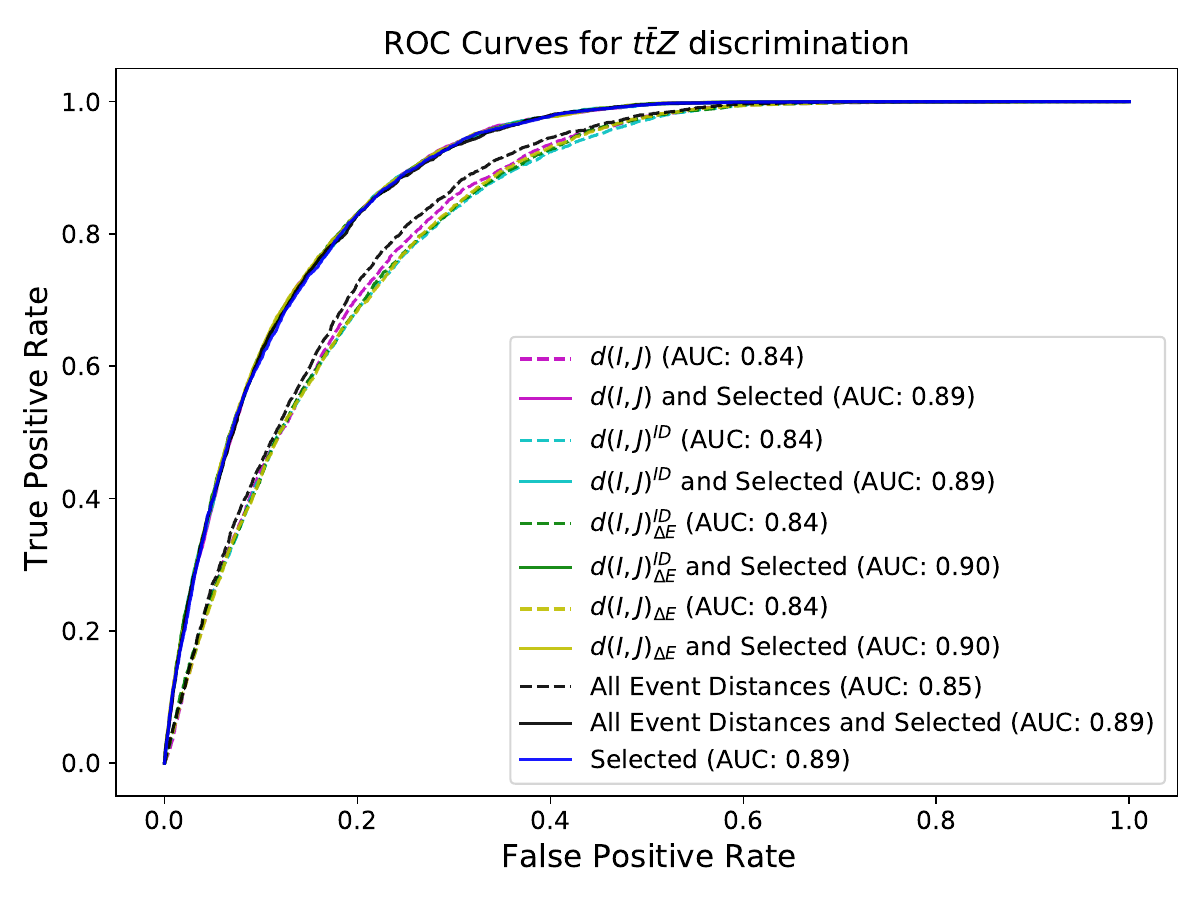}
\end{figure}

Finally, in~\cref{fig:roc_fakes}, we show how the ROC curves for the task of discriminating fakes and the $t\bar tZ$ signal from the remainder of the classes. In this figure, we see how different event distances provide different discriminant power for these specific cases. We also notice that the combination of all event distances without the selected features has better performance than each distance separately. Finally, we observe that, in the case of fake identification, the ROC curve for the combination of all event distances with the selected features is the outermost curve for the large portion of the operating points.

\onecolumngrid

\twocolumngrid

\onecolumngrid

\FloatBarrier

\twocolumngrid

\section{Conclusions}

In this paper, the Energy Mover's Distance concept was used to create a new set of observables that could be used in the measurement of rare processes at proton-proton colliders, using $t\bar{t}Z$ as a study case. We have shown that such new observables, which build on the previously proposed concept of EMD, perform well in the task of grouping together different processes based on their topologies, showing a fair discrimination power by themselves. Namely, it can be seen that the distances between $t\bar{t}Z$ and $t\bar{t}Y$ are smaller than $ZZ$ and $WZ$. This indicates that the EMD based observables can be useful in the classification of collider data.

Additionally, the use of these observables in the training of a DNN was tested. Even if the overall performance of the DNN is not, in general, significantly increased, such observables are interesting on themselves since they provide event-level information which is beneficial for the classification of processes with fake leptons in some scenarios. Furthermore, such event-level observables might be affected differently by systematic uncertainties - a study beyond the scope of the current paper which deserves further investigation.

\FloatBarrier

\section{Acknowledgments}
We thank Jesse Thaler and Guilherme Guedes for the very useful discussions. We also acknowledge the support from FCT Portugal, Lisboa2020, Compete2020, Portugal2020 and FEDER
under the projects PTDC/FIS-PAR/29147/2017 and CERN/FIS-PAR/0024/2019 and through the grant PD/BD/135435/2017.
The computational part of this work was supported by INCD (funded by FCT and FEDER under the project 01/SAICT/2016 nr. 022153) and by the Minho Advanced Computing Center (MACC).
The Titan Xp GPU card used for the training of the Deep Neural Networks developed for this project was kindly donated by the NVIDIA Corporation.

\bibliography{paper}{}
\bibliographystyle{unsrt}

\end{document}